\documentstyle[12pt,thmsa,sw20lart]{article}
\input{tcilatex}
\begin{document}

\title{{\bf Aging at the edge of chaos: Glassy dynamics and nonextensive statistics}}
\author{A. Robledo \\
%EndAName
Instituto de F\'{i}sica,\\
Universidad Nacional Aut\'{o}noma de M\'{e}xico,\\
Apartado Postal 20-364, M\'{e}xico 01000 D.F., Mexico.}
\date{.}
\maketitle

\begin{abstract}
We go over our finding that the dynamics at the noise-perturbed edge of
chaos in logistic maps is comparable to that observed in supercooled liquids
close to vitrification. That is, the three major features of glassy dynamics
in structural glass formers, two-step relaxation, aging, and a relationship
between relaxation time and configurational entropy, are displayed by orbits
with vanishing Lyapunov exponent. The known properties in control-parameter
space of the noise-induced bifurcation gap play a central role in
determining the characteristics of dynamical relaxation at the chaos
threshold. Time evolution is obtained from the Feigenbaum RG transformation,
it is expressed analytically via $q$-exponentials, and described in terms of
nonextensive statistics.\medskip


\end{abstract}

\section{Introduction}

The problem of understanding the dynamics of glass formation is known to be
difficult and challenging and therefore continues to be a frontier topic in
statistical physics \cite{debenedetti1}. It is generally thought that
ordinary phase-space mixing is not wholly realized during glass forming
dynamics, i.e. upon cooling, caged molecules rearrange so slowly that they
cannot sample configurations in the available time allowed by the process 
\cite{debenedetti1}. The advance to the glass transition is signaled by very
pronounced deceleration of relaxation processes since the characteristic
times for their completion increase several orders of magnitude within a
small interval of temperatures. Because of this extreme circumstances an
important theoretical issue is to determine whether under conditions of
ergodicity malfunction, and, as a final point, downright failure, the
Boltzmann-Gibbs (BG)\ statistical mechanics is still capable of describing
stationary states on the point of glass formation or those representing the
glass itself. This question and also that about the possible applicability
of generalizations of the BG statistics, such as the nonextensive statistics 
\cite{tsallis0} \cite{tsallis1}, to glass formation is a subject of current
concern.

As a fresh approach we study the dynamics of a model in which only the
essential ergodicity breakdown feature is present and where there is neither
molecular structure nor molecular interactions. The general aim is to learn
if the occurrence of the key properties of glassy dynamics are due primarily
to the approach to a nonergodic state and therefore are observable in
completely different types of systems, hence suggesting some degree of
universality. Our minimal model for glass dynamics is the prototypical
logistic map at the edge of chaos but perturbed by small amplitude noise. It
has only one degree of freedom but the consideration of external noise could
be thought to represent the effect of many other systems coupled to it, like
in the so-called coupled map lattices \cite{kaneko1}. Recently \cite
{robledo0}, we have proved that the dynamics of this map shows robust
similarities with that of supercooled liquids close to vitrification. This
is so since two-step relaxation, aging, and a relationship between
relaxation time and configurational entropy, are the three main
(phenomenologically established) characteristics of glassy dynamics in
structural glass formers \cite{debenedetti1}. The basic ingredient of
ergodicity failure is obtained for orbits at the onset of chaos in the limit
towards vanishing noise amplitude. Here expand the discussion of our results 
\cite{robledo0}.

Our view is that glass formation is one in which the system is driven
gradually into a nonergodic state by reducing its ability to pass through
phase-space-filling configurational regions until it is only possible to go
across a (multi)fractal subset of phase space. This situation is generated
in the logistic map with additive external noise, 
\begin{equation}
x_{t+1}=f_{\mu }(x_{t})=1-\mu x_{t}^{2}+\chi _{t}\sigma ,\;-1\leq x_{t}\leq
1,0\leq \mu \leq 2,  \label{logistic1}
\end{equation}
where $\chi _{t}$ is Gaussian-distributed with average $\left\langle \chi
_{t}\chi _{t^{\prime }}\right\rangle =\delta _{t.t^{\prime }}$, and $\sigma $
measures the noise intensity \cite{schuster1} \cite{crutchfield1}. Notice
that the formula (\ref{logistic1}) can be written as a discrete form for a
Langevin equation.

\section{Dynamics of glass formation}

We recall the main dynamical properties displayed by supercooled liquids on
approach to glass formation. One is the growth of a plateau and for that
reason a two-step process of relaxation, as presented by the time evolution
of correlations e.g. the intermediate scattering function $F_{k}(\Delta t)$ 
\cite{debenedetti1}. This consists of a primary power-law decay in time
difference $\Delta t$ (so-called $\beta $ relaxation) that leads into the
plateau, the duration $t_{x}$ of which diverges also as a power law of the
difference $T-T_{g}$ as the temperature $T$ decreases to a glass temperature 
$T_{g}$. An 'ideal' glass transition has been conjectured to occur at a $%
T_{c}$ which is not experimentally accessible but that approaches $T_{g}$ at
infinitely slow cooling in fragile glass formers \cite{mezard1}. After $%
t_{x} $ there is a secondary power law decay (so-called $\alpha $
relaxation) away from the plateau \cite{debenedetti1}. A second important
(nonequilibrium) dynamic property of glasses is the loss of time translation
invariance observed for $T$ below $T_{g}$, a characteristic known as aging 
\cite{bouchaud1}, that is due to the fact that properties of glasses depend
on the procedure by which they are obtained. Remarkably, the time fall off
of relaxation functions and correlations display a scaling dependence on the
ratio $t/t_{w}$ where $t_{w}$ is a waiting time. A third notable property is
that the experimentally observed relaxation behavior of supercooled liquids
is effectively described, via reasonable heat capacity assumptions \cite
{debenedetti1}, by the so-called Adam-Gibbs equation, $t_{x}=A\exp
(B/TS_{c}) $, where the relaxation time $t_{x}$ can be identified with the
viscosity or the inverse of the difussivity, and the configurational entropy 
$S_{c}$ is related to the number of minima of the fluid's potential energy
surface (and $A$ and $B$ are constants) \cite{debenedetti1}. A
first-principles derivation of this equation has not been developed at
present. As the counterpart to the Adam-Gibbs formula, we show below that
our one-dimensional map model for glassy dynamics exhibits a relationship
between the plateau duration $t_{x}$, and the entropy $S_{c}$ associated to
the iterate positions (configurations) within the largest number of
phase-space bands allowed by the bifurcation gap - the noise-induced cutoff
in the period-doubling cascade \cite{schuster1}.

Our results suggest that the properties known to be basic of glassy dynamics 
\cite{debenedetti1} are likely to manifest too in completely different
physical problems, such as in nonlinear dynamics, e.g. the mentioned coupled
map lattices \cite{mousseau1}, in critical dynamics \cite{holdsworth1}, and
other fields. This hints that new predictions might be encountered in the
studies, experimental or otherwise, of slow dynamics displayed by systems
other than liquids close to the glass transition. In relation to this,
aspects of glassy dynamics have been observed in metastable quasistationary
states in microcanonical Hamiltonian systems of $N$ classical rotors with
homogeneous long-ranged interactions. For special types of initial
conditions it has been found that both two-step relaxation \cite{latora1},
where the length of the metastable plateau diverges with infinite size $%
N\rightarrow \infty $, and aging \cite{montemurro1}, \cite{pluchino1} are
present in these systems. In our simpler one-dimensional dissipative map the
amplitude $\sigma $ plays a role parallel to $T-T_{g}$ or $T-T_{c}$ in the
supercooled liquid or $1/N$ in the system of rotors. Notice that the
equivalence between $\sigma $ and $T-T_{g}$ is not literal as $\sigma $
cannot take negative values.

\section{Dynamics at the edge of chaos}

In the absence of noise $\sigma =0$ the Feigenbaum attractor at $\mu =\mu
_{c}(0)=1.40115...$ is the accumulation point of both the period doubling
and the chaotic band splitting sequences of transitions \cite{schuster1}.
Except for a set of zero measure, all the trajectories with $\mu _{c}(0)$
and initial condition $-1\leq x_{in}\leq 1$ fall into the attractor with
fractal dimension $d_{f}=0.5338...$. These trajectories represent nonergodic
states, since as $t\rightarrow \infty $ only a Cantor set of positions is
accessible out of the total phase space $-1\leq x\leq 1$. For $\sigma >0$
the noise fluctuations wipe the sharp features of the periodic attractors as
these widen into bands similar to those in the chaotic attractors,
nevertheless there remains a well-defined transition to chaos at $\mu
_{c}(\sigma )$ where the Lyapunov exponent $\lambda _{1}$ changes sign. The
period doubling of bands ends at a finite value $2^{N(\sigma )}$ as the edge
of chaos transition is approached and then decreases at the other side of
the transition. This effect displays scaling features and is referred to as
the bifurcation gap \cite{schuster1} \cite{crutchfield1}. When $\sigma >0$
the trajectories visit sequentially a set of $2^{n}$ disjoint bands or
segments leading to a cycle, but the behavior inside each band is completely
chaotic. These trajectories represent ergodic states as the accessible
positions have a fractal dimension equal to the dimension of phase space.
Thus the removal of the noise $\sigma \rightarrow 0$ leads to an ergodic to
nonergodic transition in the map and we compare its properties with those
known for the process of vitrification of a liquid as $T\rightarrow T_{g}$.

The manner in which the attractor at the onset of chaos $\mu _{c}(0)$ is
visited in time was analyzed recently with the use of the initial condition $%
x_{in}=0$ \cite{baldovin1}. It was found that the absolute values for the
positions $x_{\tau }$ of this trajectory at time-shifted $\tau =t+1$ has an
structure consisting of subsequences with a common power-law decay of the
form $\tau ^{-1/1-q}$ with $q=1-\ln 2/\ln \alpha \simeq 0.24449$, where $%
\alpha =2.50290...$ is the Feigenbaum universal constant that measures the
period-doubling amplification of iterate positions \cite{baldovin1}. That
is, the Feigenbaum attractor can be decomposed into position subsequences
generated by the time subsequences $\tau =(2k+1)2^{n}$, each obtained by
proceeding through $n=0,1,2,...$ for a fixed value of $k=0,1,2,...$. The $%
k=0 $ subsequence can be written as $x_{t}=\exp _{2-q}(-\lambda _{q}t)$ with 
$\lambda _{q}=\ln \alpha /\ln 2$, and where $\exp _{q}(x)\equiv
[1-(q-1)x]^{1/1-q}$ is the $q$-exponential function. These properties follow
from the use of $x_{in}=0$ in the scaling relation \cite{baldovin1} 
\begin{equation}
x_{\tau }=\left| g^{^{(\tau )}}(x_{in})\right| =\tau ^{-1/1-q}\left| g(\tau
^{1/1-q}x_{in})\right| .  \label{trajectory1}
\end{equation}
In the presence of noise ($\sigma $ small) one obtains instead \cite
{robledo0} 
\begin{equation}
x_{\tau }=\tau ^{-1/1-q}\left| g(\tau ^{1/1-q}x)+\chi \sigma \tau
^{1/1-r}G_{\Lambda }(\tau ^{1/1-q}x)\right| ,  \label{trajectory3}
\end{equation}
where $G_{\Lambda }(x)$ is the first order perturbation eigenfunction, and
where $r=1-\ln 2/\ln \kappa \simeq 0.6332$. Use of $x_{in}=0$ yields $%
x_{\tau }=\tau ^{-1/1-q}\left| 1+\chi \sigma \tau ^{1/1-r}\right| $ or $%
x_{t}=\exp _{2-q}(-\lambda _{q}t)\left[ 1+\chi \sigma \exp _{r}(\lambda
_{r}t)\right] $ where $t=\tau -1$ and $\lambda _{r}=\ln \kappa /\ln 2$.

At each noise level $\sigma $ there is a 'crossover' or 'relaxation' time $%
t_{x}=\tau _{x}-1$ when the fluctuations start suppressing the fine
structure of the orbits with $x_{in}=0$. This time is given by $\tau
_{x}=\sigma ^{r-1}$, the time when the fluctuation term in the perturbation
expression for $x_{\tau }$ becomes unbounded by $\sigma $, i.e. $x_{\tau
_{x}}=\tau _{x}^{-1/1-q}\left| 1+\chi \right| $. There are two regimes for
time evolution at $\mu _{c}(\sigma )$. When $\tau <\tau _{x}$ the
fluctuations are smaller than the distances between neighboring subsequence
positions of the $\sigma =0$ orbit at $\mu _{c}(0)$, and the iterate
positions with $\sigma >0$ fall within small non overlapping bands each
around the $\sigma =0$ position for that $\tau $. Time evolution follows a
subsequence pattern analogous to that in the noiseless case. When $\tau \sim
\tau _{x}$ the width of the noise-generated band reached at time $\tau
_{x}=2^{N}$ matches the distance between adjacent positions where $N\sim
-\ln \sigma /\ln \kappa $, and this implies a cutoff in the progress along
the position subsequences. At longer times $\tau >\tau _{x}$ the orbits no
longer follow the detailed period-doubling structure of the attractor. The
iterates now trail through increasingly chaotic trajectories as bands merge
with time. This is the dynamical image - observed along the time evolution
for the orbits of a single state $\mu _{c}(\sigma )$ - of the static
bifurcation gap originally described in terms of the variation of the
control parameter $\mu $ \cite{crutchfield1}, \cite{crutchfield2}, \cite
{shraiman1}. The plateau structure of relaxation and the crossover time $%
t_{x}$ can be clearly observed in Fig. 1b in Ref. \cite{baldovin2} where $%
<x_{t}^{2}>-$ $<x_{t}>^{2}$ is shown for several values of $\sigma $.

\section{Parallels with glassy dynamics}

At noise level $\sigma $ the orbits visit points within the set of $2^{N}$
bands and, as explained in Ref. \cite{robledo0}, this takes place in time in
the same way that period doubling and band merging proceeds in the presence
of a bifurcation gap when the control parameter is run through the interval $%
0\leq \mu \leq 2$. Namely, the trajectories starting at $x_{in}=0$ duplicate
the number of visited bands at times $\tau =2^{n}$, $n=1,...,N$, the
bifurcation gap is reached at $\tau _{x}=$ $2^{N}$, after which the orbits
fall within bands that merge by pairs at times $\tau =2^{N+n}$, $n=1,...,N$.
The sensitivity to initial conditions grows as $\xi _{t}=\exp _{q}(\lambda
_{q}t)$ ($q=1-\ln 2/\ln \alpha <1$) for $t<t_{x}$, but for $t>t_{x}$ the
fluctuations dominate and $\xi _{t}$ grows exponentially as the trajectory
has become chaotic ($q=1$) \cite{robledo0}. This behavior was interpreted 
\cite{robledo0} to be the dynamical system analog of the $\alpha $
relaxation in supercooled fluids. The plateau duration $t_{x}\rightarrow
\infty $ as $\sigma \rightarrow 0$. Additionally, trajectories with initial
conditions $x_{in}$ not belonging to the attractor exhibit an initial
relaxation process towards the plateau as the orbit approaches the
attractor. This is the map analog of the $\beta $ relaxation in supercooled
liquids.

The entropy $S_{c}(\mu _{c}(\sigma ))$ associated to the distribution of
iterate positions (configurations) within the set of $2^{N}$ bands was
determined in Ref. \cite{robledo0}. This entropy has the form $S_{c}(\mu
_{c}(\sigma ))=2^{N}\sigma s$, since each of the $2^{N}$ bands contributes
with an entropy $\sigma s$, where $s=-\int_{-1}^{1}p(\chi )\ln p(\chi )d\chi 
$ and where $p(\chi )$ is the distribution for the noise random variable.
Given that $2^{N}=$ $1+t_{x}$ and $\sigma =(1+t_{x})^{-1/1-r}$, one has $%
S_{c}(\mu _{c},t_{x})/s=(1+t_{x})^{-r/1-r}$or, conversely, 
\begin{equation}
t_{x}=(s/S_{c})^{(1-r)/r}.  \label{adamgibbs1}
\end{equation}
Since $t_{x}\simeq \sigma ^{r-1}$, $r-1\simeq -0.3668$ and $(1-r)/r\simeq
0.5792$ then $t_{x}\rightarrow \infty $ and $S_{c}\rightarrow 0$ as $\sigma
\rightarrow 0$, i.e. the relaxation time diverges as the 'landscape' entropy
vanishes. We interpret this relationship between $t_{x}$ and the entropy $%
S_{c}$ to be the dynamical system analog of the Adam-Gibbs formula for a
supercooled liquid. Notice that Eq.(\ref{adamgibbs1}) is a power law in $%
S_{c}^{-1}$ while for structural glasses it is an exponential in $S_{c}^{-1}$
\cite{debenedetti1}. This difference is significant as it indicates how the
superposition of molecular structure and dynamics upon the bare ergodicity
breakdown phenomenon built in the map modifies the vitrification properties.

The aging scaling property of the trajectories $x_{t}$ at $\mu _{c}(\sigma )$
was examined in Ref. \cite{robledo0}. The case $\sigma =0$ is readily
understood because this property is actually built into the position
subsequences $x_{\tau }=\left| g^{(\tau )}(0)\right| $, $\tau =(2k+1)2^{n}$, 
$k,n=0,1,...$ referred to above. These subsequences can be employed for the
description of trajectories that are at first held at a given attractor
position for a waiting period of time $t_{w}$ and then released to the
normal iterative procedure. For illustrative purposes we select the holding
positions to be any of those for a waiting time $t_{w}=2k+1$, $k=0,1,...$
and notice that for the $x_{in}=0$ orbit these positions are visited at odd
iteration times. The lower-bound positions for these trajectories are given
by those of the subsequences at times $(2k+1)2^{n}$. See Fig. 1 in Ref. \cite
{robledo0}. Writing $\tau $ as $\tau =$ $t_{w}+t$ we have that $%
t/t_{w}=2^{n}-1$ and $x_{t+t_{w}}=g^{(t_{w})}(0)g^{(t/t_{w})}(0)$ or 
\begin{equation}
x_{t+t_{w}}=g^{(t_{w})}(0)\exp _{q}(-\lambda _{q}t/t_{w}).
\label{trajectory5}
\end{equation}
This fully developed aging property is gradually modified when noise is
turned on. The presence of a bifurcation gap limits its range of validity to
total times $t_{w}+t$ $<t_{x}(\sigma )$ and so progressively disappears as $%
\sigma $ is increased. Recently \cite{barkai1} aging has been observed in
the subdiffusion associated to a nonlinear map that, as it is the case here,
has $\lambda _{1}=0$ \cite{baldovin3}.

\section{Discussion}

Thus, the dynamics of noise-perturbed logistic maps at the chaos threshold
exhibits the most prominent features of glassy dynamics in supercooled
liquids. The existence of this analogy cannot be considered accidental since
the limit of vanishing noise amplitude $\sigma \rightarrow 0$ (the
counterpart of the limit $T-T_{g}\rightarrow 0$ in the supercooled liquid)
entails loss of ergodicity. The incidence of these properties in such simple
dynamical systems, with only a few degrees of freedom and no reference to
molecular interactions, suggests a universal mechanism underlying the
dynamics of glass formation. As definitely proved \cite{baldovin1}, the
dynamics of deterministic unimodal maps at the edge of chaos is a genuine
example of the pertinence of nonextensive statistics in describing states
with vanishing ordinary Lyapunov exponent $\lambda _{1}$. Here we have shown
that this nonergodic state corresponds to the limiting state, $\sigma
\rightarrow 0$, $t_{x}\rightarrow \infty $, for a family of small $\sigma $
noisy states with glassy properties, that are noticeably described for $%
t<t_{x}$ via the $q$-exponentials of the nonextensive formalism \cite
{baldovin1}.

To ascertain the degree of parallelism between the map at $\mu _{c}(\sigma )$
and a thermal system we keep in mind the following criteria: The exponential 
$\xi _{t}$ (or $\lambda _{1}>0$) of a chaotic state is the counterpart of
the BG equilibrium state. On the other hand, a power-law $\xi _{t}$ (or $%
\lambda _{1}=0$) of an incipient chaotic state represents an ''anomalous
stationary state'', $q\neq 1$, and in the case of $\mu _{c}(0)$ a nonergodic
state. Only when $\sigma =0$ we have a true anomalous stationary state (of
infinite duration) that exhibits a precise aging property. In the map at $%
\mu _{c}(\sigma )$ we find two regimes separated by $t_{x}$ with presence of
aging when $t<t_{x}$. In glass formers the two-step relaxation is seen in
equilibrium two-time correlations while aging is a manifestation of the
system having fallen out of equilibrium. Because of these apparent
differences a more detailed comparison between the map and the thermal
system properties is required, and it may be necessary to consider
neighbouring states around $\mu _{c}(\sigma )$. These states would be
expected to display also slow dynamics. When $\mu >\mu _{c}(\sigma )$ one
has $\lambda _{1}>0$ and the analog of BG equilibrium, whereas for $\mu <\mu
_{c}(\sigma )$ one has $\lambda _{1}<0$ and a situation ''out of
equilibrium''.

The anomalous stationary state has properties akin to both equilibrium and
out of equilibrium. Equilibrium-like since it is stationary, an orbit that
starts in the attractor remains there forever. Nonequilibrium-like because
the {\it finite-time} Lyapunov exponents fluctuate and take negative values
(implying confinement in phase space). When $\sigma >0$ it would be expected
that the frequency of these negative values decreases sharply after $t_{x}$.
Further studies are needed here. Also, there remains the question about
whether some features of glassy dynamics, such as loss of time translation
invariance, are due to nonergodicity or to merely being out of equilibrium
(i.e. in a non-stationary state).

It has been suggested on several occasions \cite{tsallis1} that the setting
in which nonextensive statistics appears to emerge is linked to the
incidence of nonuniform convergence, such as that involving the
thermodynamic $N\rightarrow \infty $ and very large time $t\rightarrow
\infty $ limits. For example, in the rotor problem mentioned above - for
specific choices of initial conditions - if $N\rightarrow \infty $ is taken
before $t\rightarrow \infty $ the anomalous metastable states with
noncanonical properties appear to be the only observable stationary states,
whereas if $t\rightarrow \infty $ is taken before $N\rightarrow \infty $ the
usual BG equilibrium states are obtained. Here it is clear that a similar
situation takes place, that is, if $\sigma \rightarrow 0$ is taken before $%
t\rightarrow \infty $ a nonergodic orbit confined to the Feigenbaum
attractor is obtained, whereas if $t\rightarrow \infty $ is taken before $%
\sigma \rightarrow 0$ a typical $q=1$ chaotic (ergodic) orbit is observed.

Finally, it is worth mentioning that while the properties displayed by the
map capture in a qualitative, heuristic, way the phenomenological issues of
vitrification, they are obtained in a quantitative and rigorous manner as
the map is concerned. Our map setup is a rarely available 'laboratory' where
every aspect of the dynamics can be studied analytically.

Acknowledgments. I am grateful to Piero Tartaglia for informative
discussions and also for his kind hospitality at Dipartamento di Fisica,
Universit\'{a} degli Studi di Roma ''La Sapienza''. I thank an anonymous
referee for important criticism. Work partially supported by CONACyT grant
P-40530-F.

\medskip

\end{document}